\documentclass[twocolumn,amsmath,amssymb,prl]{revtex4}
\usepackage{latexsym}
\usepackage{graphicx}
\usepackage{siunitx}
\usepackage{xcolor}
\usepackage{soul}
\usepackage[capitalise]{cleveref}
\setcitestyle{super,open={},close={}}

\makeatletter
\renewcommand\@biblabel[1]{#1.}
\makeatother

\begin{document}

\preprint{}

\title{Evolution of ${4\pi}$-periodic Supercurrent in the Presence of In-plane Magnetic Field}

\author{Bassel Heiba Elfeky$^{1}$}

\author{Joseph J. Cuozzo$^{2}$}

\author{Neda Lotfizadeh$^{1}$}

\author{William F. Schiela$^{1}$}

\author{Seyed M. Farzaneh$^{1}$}

\author{William M. Strickland$^{1}$}

\author{Dylan Langone$^{1}$}

\author{Enrico Rossi$^{2}$}

\author{Javad Shabani$^{1}$}
\email{jshabani@nyu.edu}

\affiliation{$^{1}$Center for Quantum Information Physics, Department of Physics, New York University, NY 10003, USA}

\affiliation{$^{2}$Department of Physics, William $\&$ Mary, Williamsburg, VA, 23187, USA}

\date{\today}


\begin{abstract}

\textbf{
  In the presence of a 4$\pi$-periodic contribution to the current phase relation, for example in topological Josephson junctions, odd Shapiro steps are expected to be missing. While missing odd Shapiro steps have been observed in several material systems and interpreted in the context of topological superconductivity, they have also been observed in topologically trivial junctions.
  Here, we study the evolution of such trivial missing odd Shapiro steps in Al-InAs junctions in the presence of an in-plane magnetic field $B^{\theta}$.
  We find that the odd steps reappear at a crossover $B^{\theta}$ value, exhibiting an in-plane field angle anisotropy that depends on spin-orbit coupling effects. We interpret this behavior by theoretically analyzing the Andreev bound state spectrum and the transitions induced by the non-adiabatic dynamics of the junction and attribute the observed anisotropy to mode-to-mode coupling. 
  %
  %
  %
Our results highlight the complex phenomenology of missing Shapiro steps and the underlying current phase relations in planar Josephson junctions designed to realize Majorana states.
}
\end{abstract}

\maketitle

Josephson junctions (JJs) fabricated on semiconductor structures with epitaxially grown superconductors have recently attracted attention due to their propitious characteristics \cite{krogstrup_epitaxy_2015, shabani_two-dimensional_2016, kjaergaard_transparent_2017, bottcher_superconducting_2018, mayer_superconducting_2019, lee_transport_2019, mayer_gate_2020, elfeky_local_2021, strickland_controlling_2022} and applications in quantum computing \cite{Larsen_PRL, Luthi2018, kringhoj2018, casparis_superconducting_2018, CasparisBenchmarking, yuan2021, danilenko2022, hertel2022, strickland_superconducting_2022}. In the presence of a Zeeman field \cite{hell_two-dimensional_2017, pientka_topological_2017, dartiailh_phase_2021} or a phase bias \cite{ren_topological_2019, fornieri_evidence_2019, banerjee_signatures_2022}, and a strong spin-orbit coupling (SOC) interaction, such high-quality JJs have shown signatures of topological superconductivity \cite{ren_topological_2019, fornieri_evidence_2019, dartiailh_phase_2021, banerjee_signatures_2022}, which can host Majorana zero modes useful for fault-tolerant quantum computation \cite{nayak_non-abelian_2008, aasen_milestones_2016}. However, robust implementation and signatures of topological superconductivity remain ambiguous \cite{ mohapatra_observation_2019,pan_generic_2020, cayao_confinement-induced_2021,yu_non-majorana_2021, pan_-demand_2022}.

To harness the potential of topological superconductivity, it is essential to be able to identify unambiguously
the topological character of the states in a JJ.
Topological JJs exhibit a unique fractional Josephson effect which is inaccessible with DC measurements due to relaxation processes to the ground state.
%
Consequently, detecting the fractional Josephson effect requires measurements on timescales shorter than the relaxation time
\cite{kwon_fractional_2003, shaw_kinetics_2008, martinis_energy_2009, pikulin_phenomenology_2012, badiane_ac_2013, van_woerkom_one_2015};
timescales that are accessible using microwave excitations
\cite{rokhinson_fractional_2012, wiedenmann_4-periodic_2016, bocquillon_gapless_2017, deacon_josephson_2017, laroche_observation_2019}.

\begin{figure*}[ht!]
    \centerline{\includegraphics[width=1.0\textwidth]{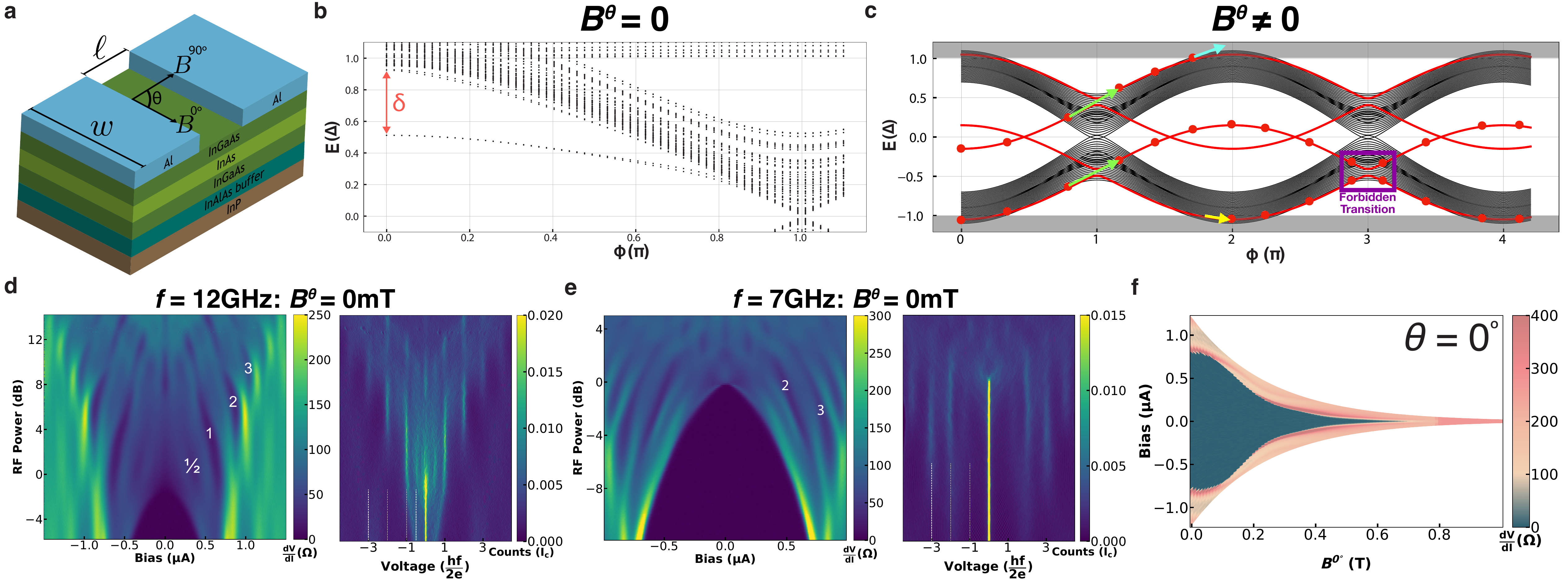}}
    \caption{\textbf{Josephson junction geometry, Andreev spectrum and characterization. a} Schematic drawing of the material heterostructure with a junction of width $w$ and length $l$ made of Al superconducting contacts and an InAs surface quantum well. The 2D axis represents the direction of an applied in-plane magnetic field, where $\theta$ is the in-plane field angle such that $B^{0^{\circ}}$ is the in-plane field along the junction and $B^{90^{\circ}}$ is the in-plane field along the current.
    \textbf{b} Example of calculated energy spectrum of the Andreev bound states in a wide junction with no applied magnetic field. The results obtained are for a JJ with $w = \SI{500}{\nano m}$ and $l = \SI{100}{\nano m}$, superconducting gap $\Delta=\SI{300}{\micro eV}$, and carrier density $n=4 \times 10^{11}\SI{}{\centi m}^{-2}$. The long junction modes that contribute to $I_{4\pi}$ are separated by $\delta$ from the quasicontinuum at $E\sim\Delta$.
    \textbf{c} Energy spectrum in the presence of a finite magnetic field with the spin-split long junction modes (red). The dots on the modes indicate an occupied state. Arrows indicate possible mode to mode (green) and mode to continuum (light-blue/yellow) transitions. 
    The dark-blue arrow indicates relaxation processes that fill low energy unoccupied states.
    The purple box defines a forbidden transition due to both states being occupied.
    \textbf{d}, \textbf{e} Differential resistance as a function of current bias, along with a histogram of the voltage distribution, all as a function of RF power, for frequency: \textbf{d} $f = \SI{12}{\giga Hz}$, and \textbf{e} $f = \SI{7}{\giga Hz}$ with no applied field, $B^{\theta} =\SI{0}{\milli T}$, for JJ1. The numbers correspond to the index of the Shapiro steps.
    \textbf{f} Differential resistance as a function of current bias and $B^{0^{\circ}}$ for JJ1.}
    \label{fig:fig_intro}
\end{figure*}

When a microwave bias is applied to a JJ, the periodic modulation of the current bias
becomes phase locked with the dynamics of the junction and results in constant voltage steps in the voltage-current characteristic known as Shapiro steps. The Andreev bound states (ABSs) of a conventional JJ in the short ballistic regime are 2$\pi$-periodic in phase $\phi$, resulting in Shapiro steps at values of $n\frac{hf}{2e}$, where $f$ is the frequency of the microwave drive, and $n$ is an integer. When the current phase relation (CPR) is 4$\pi$-periodic, as expected for a topological JJs, the fractional Josephson effect results in Shapiro steps only at $n\frac{hf}{e}$, resulting in missing odd Shapiro steps. Missing Shapiro steps have been observed in different material systems and are usually attributed to the presence of a topological state  \cite{rokhinson_fractional_2012, wiedenmann_4-periodic_2016,bocquillon_gapless_2017, li_4-periodic_2018, laroche_observation_2019, fischer_4_2022}.
In practice, even for a topological JJ, a 4$\pi$-periodic component CPR coexists with a 2$\pi$-periodic component
in which case the absence of odd Shapiro steps depends on the details of the junction, and the frequency and power of the
microwave radiation \cite{dominguez_dynamical_2012, dominguez_josephson_2017, pico-cortes_signatures_2017}.

Recent work \cite{dartiailh_missing_2021} has experimentally shown that topologically trivial JJs can also exhibit
missing odd Shapiro steps as predicted previously by other theoretical works \cite{billangeon_ac_2007,dominguez_dynamical_2012, sau_detecting_2017, galaktionov_fractional_2021,fischer_4_2022}. This can happen
when ABSs with a large probability of undergoing a Landau-Zener transition (LZT) at $\phi \sim \pi$,
and a negligible probability of crossing into the continuum, are present.
Other mechanisms responsible for missing Shapiro steps have also been proposed involving a bias-dependent junction resistance \cite{mudi_model_2021}, or the presence of multiband superconducting states \cite{Cuozzo2022}. Therefore, the observation of 4$\pi$-periodic supercurrent I$_{4\pi}$ or missing Shapiro steps is a necessary signature of topological superconductivity but is not conclusive.
Given that an in-plane magnetic field $B^\theta$ is one of the ingredients required to drive a JJ to a topological transition, understanding how missing Shapiro steps depend on $B^\theta$ is essential to distinguish a trivial JJ from its topological counterpart.

In this work, we present measurements on highly-transparent epitaxial Al-InAs JJs in the presence of an in-plane magnetic field $B^{\theta}$
and SOC effects, conditions associated with inducing topological superconductivity. 
For $B^{\theta}=\SI{0}{\milli T}$, we observe missing odd Shapiro steps with no applied field due to the presence of a topologically trivial $I_{4\pi}$ as observed previously \cite{dartiailh_missing_2021}. As $B^{\theta}$ is increased, these missing Shapiro steps eventually reappear and no topological signatures are observed up to the junction critical field $B^{\theta}_{c}$. The reappearance of the missing steps exhibits angle anisotropies that depend on the angle-dependent $B^{\theta}_{c}$ and carrier density associated with SOC interaction effects. Our results show the complex dependence of topologically trivial $I_{4\pi}$ on the applied in-plane field magnitude and direction, and SOC effects.



\begin{figure*}[ht!]
    \centerline{\includegraphics[width=1.0\textwidth]{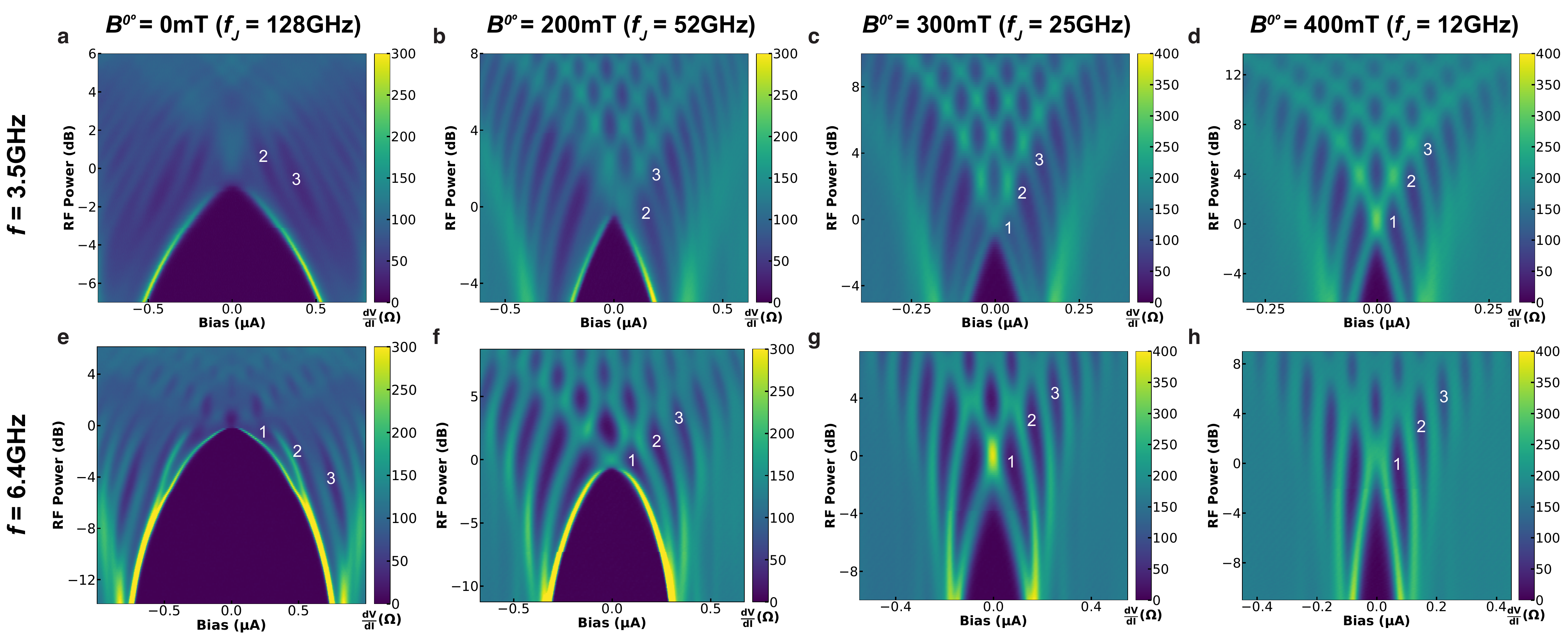}}
    \caption{\textbf{Missing Shapiro step reemergence at finite in-plane magnetic field.}
    Differential resistance as a function of current bias and RF power at \textbf{a-d} $f = \SI{3.5}{\giga Hz}$ and \textbf{e-h} $f = \SI{6.4}{\giga Hz}$ for different $B^{0^{\circ}}$ values for JJ1.
    }
    \label{fig:fig_shapiro_field_freq}
\end{figure*}

\cref{fig:fig_intro}a presents the junction heterostructure studied. An InAs near-surface quantum well is grown between two layers of In$_{0.81}$Ga$_{0.19}$As which is then capped with a thin layer of epitaxial Al grown \textit{in situ}.
Two JJs, JJ1 and JJ2, are fabricated on two different wafers grown under slightly different growth conditions (see Supporting Information). The junctions are defined using a selective wet etch of the Al and are $w=\SI{4}{\micro m}$ wide and $l\sim\SI{100}{\nano m}$ long. 
Given $l$ of the junctions, the calculated mean free path to be $l_{\rm mfp} \approx 150-\SI{250}{\nano m}$, 
and the superconducting
coherence length $\xi\approx 530-\SI{630}{\nano m}$, the junctions are expected to be in the short ($l<\xi$) ballistic ($l<l_{\rm mfp}$) regime.

To get insight into the dynamics of such highly transparent junctions, we first perform tight binding simulations of an Al-InAs junction using realistic parameters and calculate the energy spectrum of the ABSs shown in \cref{fig:fig_intro}b (simulation details are provided in Supporting Information).
The calculations of these wide junctions present a complex ABS spectrum with hundreds of modes.
For a junction with width larger than the coherence length ($w > \xi$), modes with momentum primarily along the transverse direction behave effectively as ``long junction'' modes \cite{dartiailh_missing_2021}. Consequently, these modes develop a detachment gap $\delta$ from the continuum when the phase difference across the junction $\phi$ is zero, as indicated in \cref{fig:fig_intro}b.
The number of long junction modes and their $\delta$ is sensitive to several factors (density $n$, $w$,...). 
When the junction is highly transparent, the gap at $\phi = \pi$ is sufficiently small to allow Landau-Zener transitions (LZTs) when the system is diabatically driven \cite{billangeon_ac_2007, dominguez_dynamical_2012, dartiailh_missing_2021}. The combination of a large detachment gap and a small gap at $\phi=\pi$ for these long junction modes gives rise to a $4\pi$-periodic contribution to the CPR, causing a topologically trivial junction to have both $2\pi$- and $4\pi$-periodic supercurrent channels \cite{dominguez_josephson_2017, pico-cortes_signatures_2017}. In the presence of a magnetic field in the plane of the junction, the Zeeman effect splits the ABSs and eventually leads to the closing of the detachment gap of the long junction modes, as seen in \cref{fig:fig_intro}c. The $4\pi$-periodic trajectory of long junction modes is then suppressed due to transitions to the continuum. Additionally, LZTs may occur between long junction modes and other modes with negligible detachments gaps, leading to transitions to the continuum mediated by conventional ABSs and suppressing $I_{4\pi}$.

To experimentally investigate such trivial $4\pi$-modes, we examine the microwave response of JJ1 in a DC current-biased setup. The measurements are carried out at $T=\SI{30}{\milli K}$ where the junction exhibits no hysteresis, as seen in Supporting Fig. S2. In \cref{fig:fig_intro}d, we present $\frac{dV}{dI}$ as a function of the DC current bias and RF power at $f=\SI{12}{\giga Hz}$ in addition to a histogram of the voltage distribution. For this value of $f$, we can identify all the integer Shapiro steps along with subharmonic Shapiro steps.
Subharmonic Shapiro steps are expected at high frequencies due to the anharmoncity associated with the forward skewness of the CPR in highly transparent junctions \cite{lee_ultimately_2015, askerzade_effects_2015, wiedenmann_4-periodic_2016, snyder_weak-link_2018,kringhoj_anharmonicity_2018, panghotra_giant_2020, oconnell_yuan_epitaxial_2021}.
The presence of a $4\pi$-periodic supercurrent channel, with critical current $I_{4\pi}$, is expected to result in missing odd Shapiro steps \cite{billangeon_ac_2007,dominguez_dynamical_2012, sau_detecting_2017, dartiailh_missing_2021, galaktionov_fractional_2021, fischer_4_2022} when the energy of the photon irradiating the JJ, $hf$, is less than $hf_{4\pi} \approx 2eI_{4\pi}R_{n}$  \cite{dominguez_josephson_2017, pico-cortes_signatures_2017}. \cref{fig:fig_intro}e shows a similar Shapiro map for
$f=\SI{7}{\giga Hz}$ where we see that the first odd Shapiro step is missing indicating the presence of a finite $I_{4\pi}$ even though the JJ is in a topologically trivial regime. For JJ1, at $B^{\theta}=\SI{0}{\milli T}$, we find $f_{4\pi}\sim \SI{8.2}{\giga Hz}$
corresponding to $I_{4\pi}=\SI{52.1}{\nano A}$.
Considering the Josephson frequency, $f_{J} \equiv \frac{2e\text{I}_{c}\text{R}_{n}}{h}$,
for JJ1, we get $f_{4\pi}$/$f_{J} \cong I_{4\pi}$/$I_{c}$ corresponding to 6.5\% of the supercurrent being carried by a $4\pi$-periodic supercurrent channel.

We next consider the dependence of the critical current $I_{c}$ in JJ1 on a magnetic field, without a microwave bias, as seen in the differential resistance map in \cref{fig:fig_intro}f where the in-plane magnetic field is applied along the junction, $B^{0^{\circ}}$. The critical field, $B^{0^{\circ}}_{c}$, is seen to be $\sim\SI{620}{\milli T}$. Similar measurements performed at different $\theta$ values are presented in Supporting Fig. S3. The field dependence data show no topological signatures such as a minimum in $I_{c}$ \cite{dartiailh_phase_2021}, indicating that the junctions are topologically trivial for all the values of $B^{\theta}$ up to the critical field $B^{\theta}_{c}$.\\

\begin{figure}[ht!]
\centering
\includegraphics[width=0.95\columnwidth]{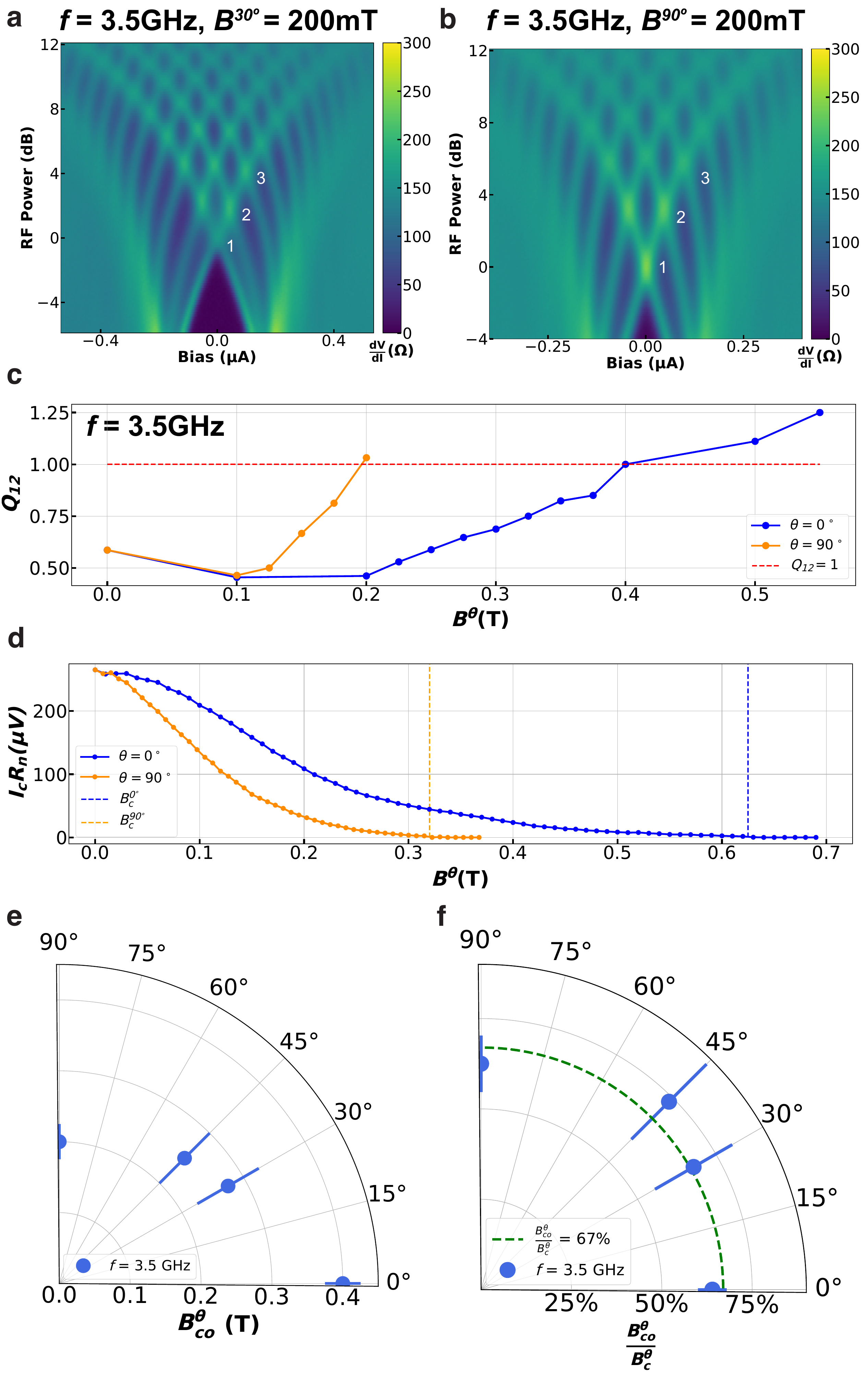}
    \caption{\textbf{Angle dependence of reemergence of missing Shapiro step.} Shapiro maps at $B^{0^{\circ}} = \SI{200}{\milli T}$ for \textbf{a} $\theta = 30^{\circ}$ and \textbf{b} $\theta = 90^{\circ}$. \textbf{c} Calculated $Q_{12}$ and \textbf{d} $I_{c}R_{n}$ as a function of in-plane magnetic field $B^{\theta}$ for in-plane field angles $\theta = 0^{\circ}$ and $90^{\circ}$.
    \textbf{e, f} The crossover field $B^{\theta}_{co}$, field value at which missing Shapiro step first fully reemerges, presented in \textbf{e} units of Tesla and \textbf{f} normalized by the corresponding critical field $B^{\theta}_{c}$, as a function of $\theta$.
}
    \label{fig:fig_shapiro_field_Q}
\end{figure}

    

In \cref{fig:fig_shapiro_field_freq}, we present Shapiro maps for various magnetic field strengths applied along the junction for $f=\SI{3.5}{\giga Hz}$ and $f=\SI{6.4}{\giga Hz}$. At $B^{0^{\circ}} = \SI{0}{\milli T}$, the first Shapiro step is seen to be missing for both frequencies since $f < f_{4\pi}$. At $B^{0^{\circ}} = \SI{200}{\milli T}$, the first step almost completely emerges for $f=\SI{6.4}{\giga Hz}$ while still being missing for $f=\SI{3.5}{\giga Hz}$. At $B^{0^{\circ}} \sim \SI{300}{\milli T}$, the first step starts emerging for $f=\SI{3.5}{\giga Hz}$, eventually completely appearing at $B^{0^{\circ}} = \SI{400}{\milli T}$. This behavior implies a decrease of $I_{4\pi}$ as a function of in-plane field strength, consistent with the mechanisms described in \cref{fig:fig_intro}c. We note that the data presented in \cref{fig:fig_shapiro_field_freq} imply that $f_{4\pi}$ does not scale proportionally with $f_{J}$. In fact, the ratio $f_{4\pi} / f_{J}$ generally increases as a function of in-plane field strength. This indicates 
that the suppression of $I_{4\pi}$ is not simply proportional to the critical current $I_c$, implying that the response of diabatically driven long junction modes to an in-plane field is distinct from conventional ``short junction'' modes that make up the rest of the spectrum in 2DEG JJs and the entire spectrum in narrow junctions e.g., nanowire junctions.


Next, we consider the $I_{4\pi}$ dependence on the applied in-plane field direction, $\theta$. A topologically non-trivial $I_{4\pi}$ is expected to be sensitive \cite{scharf_tuning_2019} to $\theta$; on the other hand, the angle dependence of a trivial $I_{4\pi}$ resulting from LZT is ambiguous and can depend on several contributing effects from Zeeman, orbital and SOC interactions. \cref{fig:fig_shapiro_field_Q}a and b show Shapiro maps with $f = \SI{3.5}{\giga Hz}$ at $B^{\theta} = \SI{200}{\milli T}$ for $\theta = 30^{\circ}$ and $\theta = 90^{\circ}$. Unlike the $\theta = 0^{\circ}$ case presented in \cref{fig:fig_shapiro_field_freq}b, the first step appears to partially reemerge for $\theta = 30^{\circ}$ and completely reemerges for $\theta = 90^{\circ}$, which indicates an angle anisotropy of $I_{4\pi}$. 
To determine more precisely the threshold value of $B^{\theta}$ above which the first step reappears, we calculate $Q_{12}$ as a function of $B^{\theta}$ where the ratio $Q_{12} = \frac{s_{1}}{s_{2}}$ represents the strength of the first step with respect to the second found by binning the voltage distribution and calculating the max step size/bin count of the first (second) step, $s_{1}$ ($s_{2}$). More details about the extraction of $Q_{12}$ from the data are provided in the Supporting Information.
We then identify the crossover field $B_{co}^\theta$ for the in-plane angle $\theta$ as the value of $B^{\theta}$ for which $Q_{12}\approx 1$.
\cref{fig:fig_shapiro_field_Q}c shows the evolution of $Q_{12}$ with $B^{\theta}$ for $\theta = 0^{\circ}$ and $\theta = 90^{\circ}$. In both cases, the first step is suppressed up to the crossover value $B_{co}^\theta$ and is fully present for values $B^{\theta}>B_{co}^\theta$.
The scaling of $Q_{12}$ is seen to exhibit clear anistropy with respect to $B^{\theta}$: $\theta =  0^{\circ}$ shows a  $B^{0^{\circ}}_{co}\approx\SI{400}{\milli T}$, whereas $\theta = 90^{\circ}$ shows a $B^{\theta}_{co}\approx\SI{200}{\milli T}$.

 \begin{figure*}[ht!]
    \centerline{\includegraphics[width=1.0\textwidth]{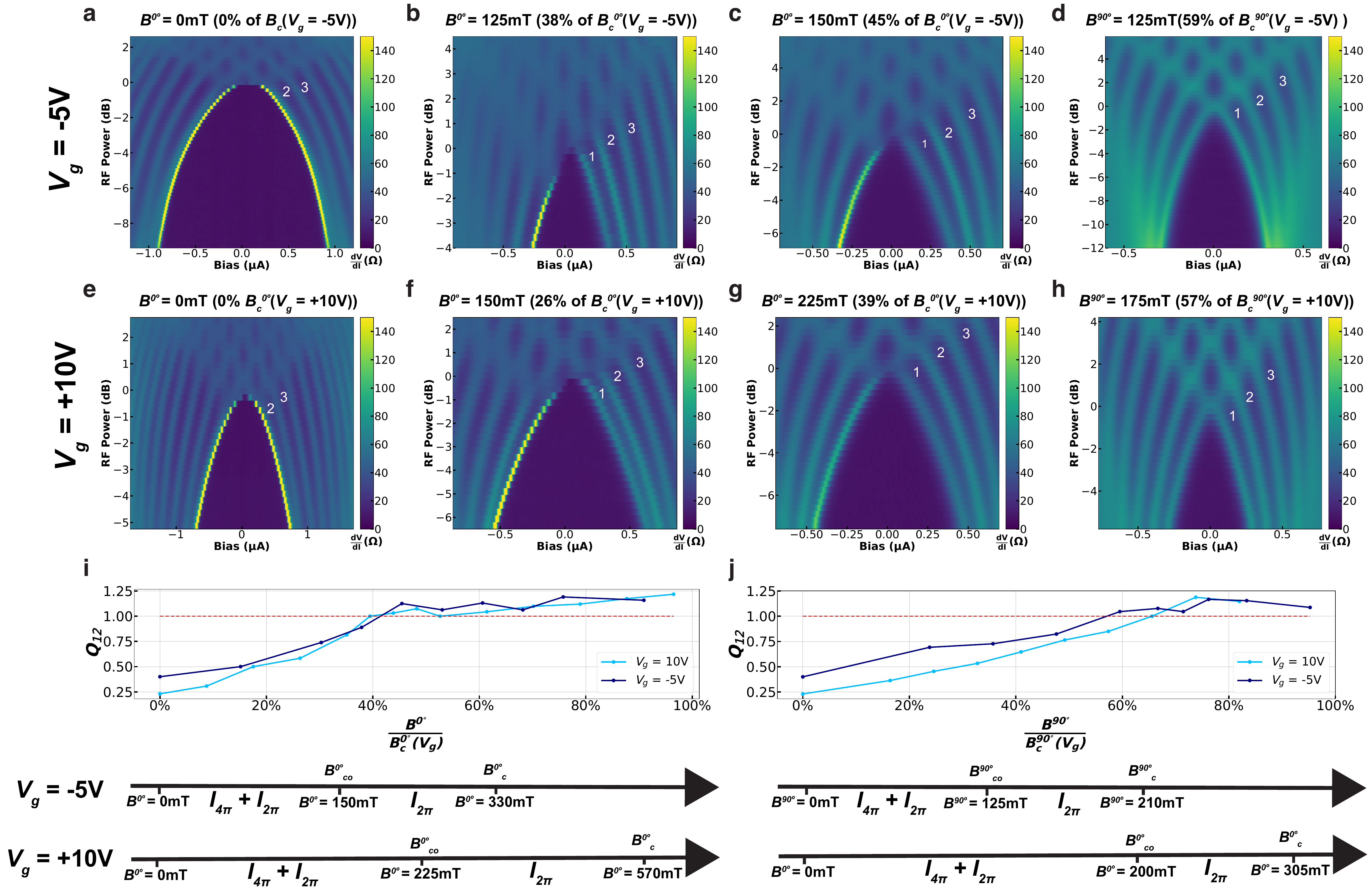}}
    \caption{\textbf{Evolution of missing Shapiro steps at an applied gate voltage for JJ2.} Shapiro maps at $f = \SI{3.4}{\giga Hz}$ for \textbf{a-d} $V_{g} = -5$V and \textbf{e-h} $V_{g} = +10$V for different $B^{0^{\circ}}$ and $B^{90^{\circ}}$ values. \textbf{i, j}$Q_{12}$ as a function of $B^{\theta}$ (normalized by the respective critical field for each $\theta$ and $V_{g}$ value) for \textbf{i} $\theta = 0^{\circ}$ and \textbf{j} $\theta = 90^{\circ}$.}
    \label{fig:fig_shapiro_field_vg}
\end{figure*}
 
In \cref{fig:fig_shapiro_field_Q}e, we present a polar plot of $B^{\theta}_{co}$ (at $f = \SI{3.5}{\giga Hz}$) as a function $\theta$. A large variation in crossover field is observed; however, we note that the critical field $B^{\theta}_{c}$ for $\theta = 0^{\circ}$ and $90^{\circ}$ are significantly different ($B^{0^{\circ}}_{c} = \SI{620}{\milli T}$ and $B^{90^{\circ}}_{c} = \SI{320}{\milli T}$) as seen in \cref{fig:fig_shapiro_field_Q}d, similar to other Al-InAs junctions \cite{suominen_anomalous_2017}. When normalized by their respective critical fields to account for the angle-dependence of $B^{\theta}_{c}$, the crossover fields become quantitatively similar and in fact match a fit of $B^{\theta}_{co}/B^{\theta}_{c}$ = 67\% as seen in \cref{fig:fig_shapiro_field_Q}f. 
This suggests that the anisotropy observed in $B^{\theta}_{co}$ is likely due to the variation in critical field and implies that JJ1 has weak SOC effects.\\ 

One of the unique advantages of using a semiconductor-based system is the ability to have electrostatic tunability of the carrier density and SOC interaction using a gate. To study the trivial $I_{4\pi}$ dependence on such properties, we focus on JJ2 fabricated on the same heterostructure presented in \cref{fig:fig_intro} but equipped with a top gate. JJ2 is expected to have a stronger SOC interaction than JJ1 even at zero gate voltage ($V_{g} = \SI{0}{V}$) due to the presence of a gate dielectric Al$_{2}$O$_{3}$ layer (see Supporting Information) that tends to increase the carrier density and consequently SOC interaction. JJ2 is markedly hysteretic at $\SI{30}{\milli K}$ due to thermal effects \cite{courtois_origin_2008}, and so it is studied at $\SI{800}{\milli K}$ where it shows no hysteresis. 
At $B^{\theta} = \SI{0}{\milli T}$ and $V_{g} = \SI{0}{V}$, JJ2 exhibits a missing first Shapiro step as seen in Supporting Fig. S8 even though at $T = \SI{800}{\milli K}$ the overall transparency is expected to be reduced. Further, Supporting Fig. S4 shows that JJ2 exhibits a similar $B^{\theta}_{c}$ anisotropy to that of JJ1. However, we note that for JJ2, $B^{\theta}_{c}$ also depends on $V_{g}$.

In \cref{fig:fig_shapiro_field_vg}, we present measurements performed on JJ2 at $f = \SI{3.4}{\giga Hz}$ for $V_{g} = -5$V and +10V at different $B^{0^{\circ}}$ and $B^{90^{\circ}}$ values. For $\theta = 0^{\circ}$, $V_{g} = -5$V shows $B^{0^{\circ}}_{co}$ = $\SI{125}{\milli T}$ while $V_{g} = +10$V shows $B^{0^{\circ}}_{co}=\SI{225}{\milli T}$. The difference between $V_{g} = -5$V and +10V is reconciled when considering $B^{0^{\circ}}/B^{0^{\circ}}_{c}(V_{g})$, as seen in \cref{fig:fig_shapiro_field_vg}i, where both $V_{g}$ values exhibit a $B^{0^{\circ}}_{co}(V_{g})/B^{0^{\circ}}_{c}(V_{g})$ of $\sim$ 40\%. 
For $\theta = 90^{\circ}$, the data presented in \cref{fig:fig_shapiro_field_vg}j show a $B^{90^{\circ}}_{co}/B^{90^{\circ}}_{c}$ ratio of $\sim57\%$ and $\sim65\%$ for $V_{g} = -5$V and +10V, respectively. While the $\theta = 90^{\circ}$ case exhibits similar $B^{90^{\circ}}_{co}/B^{90^{\circ}}_{c}$ values to that reported for JJ1, the $\theta = 0^{\circ}$ case shows a significant discrepancy for both $V_{g}$ values. It is evident here that for JJ2, the angle anisotropy is not simply accounted for by considering $B^{\theta}_{c}$ and that other effects play a role in the suppression of $I_{4\pi}$, consistent with the expectation of JJ2 having stronger SOC effects in comparison to JJ1.
In the following, we discuss the origin of such suppression of $I_{4\pi}$ and the observed angle anisotropy by considering the ABS spectrum.\\

\begin{figure*}[htbp!]
\centering
\includegraphics[width=0.99\textwidth]{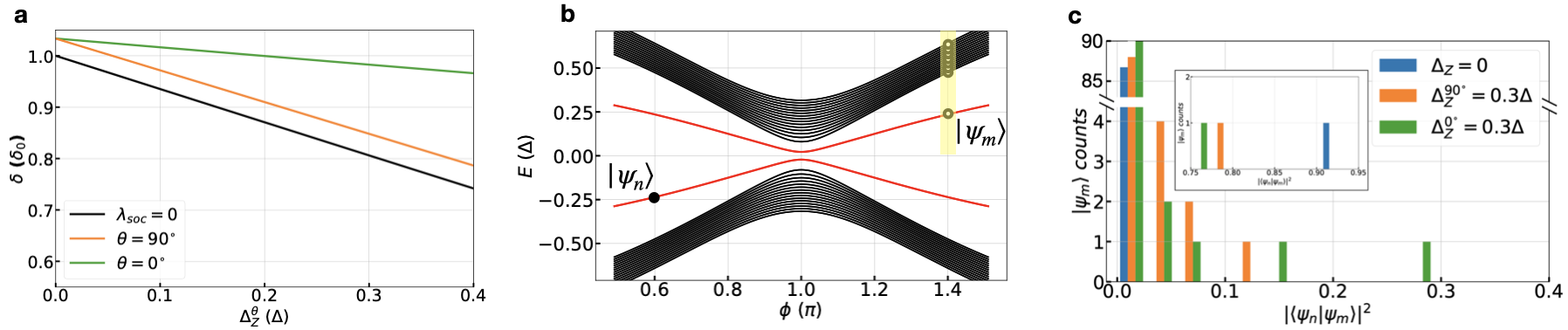}
\caption{\textbf{Theoretical ABS spectrum analysis in the presence of a Zeeman field.} \textbf{a} Calculation of the detachment gap $\delta$ for a junction with $w = \SI{500}{\nano m}$ wide and $l = \SI{100}{\nano m}$ as a function of the Zeeman energy $\Delta_Z^{\theta}$ for $\lambda_{SOC} = 0$ and for $\theta = 0^{\circ}$ and $\theta = 90^{\circ}$ with $\lambda_{SOC} = \SI{7.5}{\milli eV \cdot \nano m}$. \textbf{b} Schematic of the Andreev bound state spectrum illustrating which states are used to calculate wavefunction overlaps. \textbf{c} Distribution of wavefunction overlaps with $\lambda_{SOC} = \SI{7.5}{\milli eV \cdot \nano m}$ between $\phi_i = 0.6\pi$ and $\phi_f = 1.41\pi$ for $\Delta_Z^{\theta} = 0$ and 0.3$\Delta$ for $\theta = 0^{\circ}$ and  $90^{\circ}$. Inset: outlier overlaps where $\vert \psi_m\rangle$ is a long junction mode. 
} 
\label{fig:fig_detachgap_overlaps}
\end{figure*}

Following the picture presented in \cref{fig:fig_intro}c, we first consider the suppression of $I_{4\pi}$ in terms of transitions between the long junction modes to the continuum, related mainly to the detachment gap $\delta$. Using tight-binding simulations, we calculate the energy spectrum of the ABS spectrum in an InAs-Al junction. \cref{fig:fig_detachgap_overlaps}a shows a linear decrease in $\delta$ as a function of the Zeeman field $\Delta^{\theta}_{Z}$. The decrease in $\delta$ results in a higher probability of undergoing LZTs to the continuum, suppressing the $4\pi$-component of the CPR. In the absence of SOC effects ($\lambda_{SOC} = 0$), corresponding to the black line in \cref{fig:fig_detachgap_overlaps}a, the suppression of $\delta$ as a function of $B^{\theta}$ shows no $\theta$-dependence. 

In the presence of strong SOC effects, the Fermi surface of the quantum well has an anisotropic response to an in-plane Zeeman field, creating an anisotropic suppression of $\delta$ in the ABS spectrum. For $\lambda_{SOC} = \SI{7.5}{\milli eV \cdot \nano m}$, \cref{fig:fig_detachgap_overlaps}a illustrates that a larger $\Delta^{\theta}_{Z}$ in the $\theta = 0^{\circ}$ (green line) direction is needed than in the $\theta = 90^{\circ}$ (orange line) direction to suppress $\delta$ by the same amount. 
However, \cref{fig:fig_shapiro_field_vg}i and j shows $B^{0^{\circ}}_{co}/B^{0^{\circ}}_{c}<$ $B^{90^{\circ}}_{co}/B^{90^{\circ}}_{c}$. 
This indicates that the presence of strong SOC (as expected for JJ2) enhances the lack of correlation between the suppression of $I_{4\pi}$ and
of $\delta$.


We thus consider the suppression of $I_{4\pi}$ in terms of mode-to-mode coupling. Due to the large number of ABS modes in our junctions, a result of the large width $w$,
we have a very dense ABS spectrum.
Consequently, we have several quasi-avoided crossings between ABSs and between ABSs and the continuum.
In the presence of a Zeeman field, the ABS spectrum becomes even more complex,
with more quasi-avoided crossings and new protected crossings.
A fully microscopic description of the JJ would require the determination
of the dynamics of a multi-level Landau-Zener problem.
This is a problem that is computationally prohibitive to solve.

However, to gain a qualitative understanding, we can estimate the relevant multi-mode couplings by calculating the wave function overlap between a long junction mode at $\phi  = \phi_i$ and all positive energy Andreev mid-gap states at $\phi = \phi_f$ far from the avoided crossing at $\phi = \pi$, as shown schematically in \cref{fig:fig_detachgap_overlaps}b. 
This allows to estimate the probability that an occupied ABS, when $\phi\approx\pi$, can either transition to an ABS with a large detachment from the continuum and therefore contribute to $I_{4\pi}$, or transition
to an ABS with a small $\delta$ and therefore contribute solely to $I_{2\pi}$.
We provide a detailed discussion of the calculations in the Supporting Information. 
In \cref{fig:fig_detachgap_overlaps}c, we present a histogram of the wave function overlaps $\vert \langle \psi_n(\phi_i) \vert \psi_m(\phi_f)\rangle \vert^2$ between a long junction mode $\vert \psi_n \rangle$ and modes $\vert \psi_m \rangle$ for $\phi_i = 0.6\pi$ and $\phi_f=1.41\pi$.
%
At $\Delta_Z = 0$, we observe a distribution localized at zero except for a single outlier shown in the inset. This outlier corresponds to an overlap with another long junction mode. At finite $\Delta^{\theta}_{Z}$ and $\lambda_{SOC} = \SI{7.5}{\milli eV \cdot \nano m}$, more states develop a non-zero overlap with the long junction mode evident from the histogram distribution. The histogram distribution also shows that the system is more sensitive to $\Delta^{\theta}_{Z}$ in the $\theta=0^{\circ}$ direction than the $\theta=90^{\circ}$ direction with the $\theta=0^{\circ}$ case exhibiting a broader distribution.
%
These results suggest that the distribution of the overlaps between ABS states across $\phi=\pi$, through their effect on Landau-Zener transitions, play an important role in the anisotropy observed in \cref{fig:fig_shapiro_field_vg}i and j for JJ2, especially where a strong SOC interaction is present.

By studying the microwave response of an epitaxial Al-InAs JJ, we observe signatures of a 4$\pi$-periodic contribution to the CPR attributed to topologically-trivial LZT between long junction modes. With the application of an external magnetic field, the $I_{4\pi}$ is observed to be suppressed differently to $I_{2\pi}$ and eventually disappears at a crossover field. 
In a device with weak SOC (JJ1), we observe an isotropic suppression of $I_{4\pi}$ with an applied magnetic field when the device's angle anisotropy in $B^{\theta}_{c}$ is taken into account. In the gate tunable device (JJ2) with a significantly larger SOC, an anisotropic suppression of $I_{4\pi}$ is observed, which cannot be accounted for by the device's $B^{\theta}_{c}$ angle anisotropy. We attribute the anisotropy to SOC effects which introduce a non-trivial angle $\theta$ dependence in
the coupling of long junction modes to other Andreev mid-gap states lacking a detachment gap, suggesting multi-level LZTs. Our results indicate that such anisotropy in in-plane magnetic field and dependence on SOC effects need to be considered when differentiating between topologically trivial and non-trivial $I_{4\pi}$ and requires other correlated signatures to make claims about topological superconductivity.






{\bf Supporting Information -} Additional measurements that support our findings as well as material growth, fabrication, measurement details and information about the theoretical model are provided in the Supporting Information section. This material is available free of charge via the internet at http://pubs.acs.org.

{\bf Acknowledgments -} We thank Matthieu C. Dartiailh for fruitful discussions. The NYU team acknowledges support by DARPA TEE award no. DP18AP900007. We acknowledge funding from DOE award no. DE-SC0022245. J.J.C also acknowledges support from the Graduate Research Fellowship awarded by the Virginia Space Grant Consortium (VSGC). J.J.C. and E.R. acknowledge William \& Mary Research Computing for providing computational resources and/or technical support that have contributed to the results reported within this paper. URL: https://www.wm.edu/it/rc. W.F.S. acknowledges funding from an ND- SEG Fellowship. W.M.S. acknowledges funding from the ARO/LPS QuaCR Graduate Fellowship. 

\bibliography{Ref}

\end{document}